\documentclass{appolb}
\usepackage{graphicx}
\usepackage{subfigure}
\usepackage[T1]{fontenc}     
\usepackage[utf8]{inputenc}  
\usepackage[english]{babel}
\usepackage{bm}

\begin{document}
\title{Search
for the Charge symmetry forbidden decays of electron-positron bound state using the J-PET detector%
\thanks{Presented at the 3$^{rd}$ Jagiellonian Symposium on Fundamental and Applied Subatomic Physics}%
}
\author{J. Chhokar on behalf of J-PET Collabration
\address{Faculty of Physics, Astronomy and Applied Computer Science, Jagiellonian
University, 30-348 Cracow, Poland}
\\
}
\maketitle
\begin{abstract}
The Jagiellonian positron emission tomograph (J-PET) is a multi-purpose device built out of plastic scintillators. With large acceptance and high angular resolution, it is suitable for the studies of various phenomena such as discrete symmetries in decay of positronium atom or entangled states of photons as well as medical imaging. J-PET enables the measurement of  momenta together with photon polarization related observables. Large acceptance and high granularity of the J-PET detector enables measurement of ortho-positronium decays into three photons in the whole phase-space. In this paper we present the search of the C-forbidden decays of positronium with the J-PET detector. 
\end{abstract}
\PACS{PACS numbers come here}
  
\section{Introduction}
The word symmetry originates from ancient Greek language $\sigma\upsilon\mu\mu\epsilon\tau\rho\omega\upsilon$ which signifies well proportioned and with time symmetry involves with harmony and beauty. Early science was enormously formed by such concept of symmetry for example, the way Kepler related harmony of musical intervals to the dimension of planetary orbits. It is interesting thing to observe how modern physics has come back to reserve a very central role to this concept. Symmetries are associated with transformation of the system. If any physical system after transformation is indistinguishable from the original, we say that the system is invariant with respect to the symmetry associated to this transformation. In the beginning of the 20$^{th}$ century the celebrated Noether's theorem was proven that states if a system is invariant with respect to continuous global transformation then there exists a corresponding quantity that is conserved~\cite{Noe1918}, for example: translation in space, time or rotation conserves momentum, energy, or angular momentum of a system, respectively. Although in case of discrete symmetries such as reflection in space (P), reversal in time (T) and charge conjugation (C), Noether's theorem is not applicable. However, these symmetries appear to be conserved in the processes driven by the gravitational, electromagnetic and strong interactions up to now but were experimentally shown to be violated in weak interactions. C and CP violation in the early Universe are necessary for the occurrence of baryogenesis (for generation of asymmetry between baryons and anti-baryons) and leptogenesis~\cite{Sakh1967, Fuk86}. 
\newline 

In this article, we discuss the potential of J-PET tomograph to find forbidden  (${^1}S_0~\to~3\gamma$) and allowed (${^3}S_1~\to~3\gamma$) decay modes of the positronium (Ps)~\cite{Deut1956}, which is a bound system of lepton ($e^-$) and anti-lepton ($e^+$) with zero angular momentum in ground state, therefore, a unique tool to test the charge symmetry violation in leptonic sector ~\cite{ACTA2016, Bass18}. Charge conjugation is a transformation associated with the exchange of particles and antiparticles, by changing the sign of all additive quantum numbers (for example, electric charge). C-symmetry is violated in weak interactions and the best limit of the C symmetry violation in the electromagnetic interaction was set  with the $\pi^0 \to 3\gamma$ decays which amounts to have branching ratio (3$\gamma$/2$\gamma$) $3.1~\times~10^{-8}$ at 90$\%$ C.L.~\cite{NOW17A}. Experimental test of C-symmetry in positronium decays was performed by Mills and Berko (1967) and the best limit was set for ${^1}S_0~\to~3\gamma$ which is $2.6~\times~10^{-6}$ at 68$\%$ C.L.~\cite{MIL67A}.  According to the Standard Model predictions, photon-photon interaction or weak interaction can mimic the symmetry violation in the order of 10$^{-9}$ (photon-photon
interaction) and 10$^{-14}$ (weak interactions), respectively. The experiment measured the 3$\gamma$ rate from positronium decay, both in symmetric decay configuration (in which all three photons have the same energy and are emitted in a plane at 120$^\circ$ from each other) and in all other possible angular configurations of emitted photons in whole phase-space. To measure angular configuration of three decaying photons in the aformentioned experiment, the experimental setup consisted of only six NaI(Tl) scintillators which limited the phase space. J-PET due to its large acceptance and higher angular resolution as well as its ability to cover whole phase space allows for more precise investigation of the C-symmetry.
\newline

This article is arranged as follows: first, we explains the principle of operation of J-PET tomograph including the techniques and software which are used to analyse the experimental data. Further on, we present the production of positronium and the selection criteria for experimental data to select the allowed 3$\gamma$ decay from positronium and later we focus on the separation of allowed and forbidden 3$\gamma$ decays. 

\section{J-PET: Jagiellonian Positron Emission Tomograph}
\begin{figure}[htb]
    \begin{subfigure}{}
    \centering
    \includegraphics[width=5.0cm]{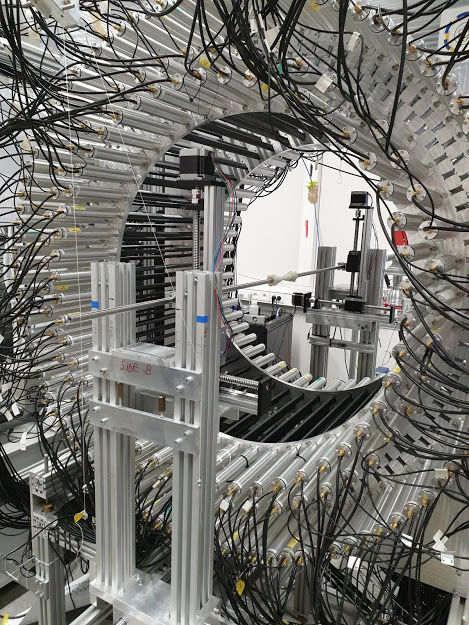}
    \end{subfigure}
    \begin{subfigure}{}
    \includegraphics[width=5.5cm]{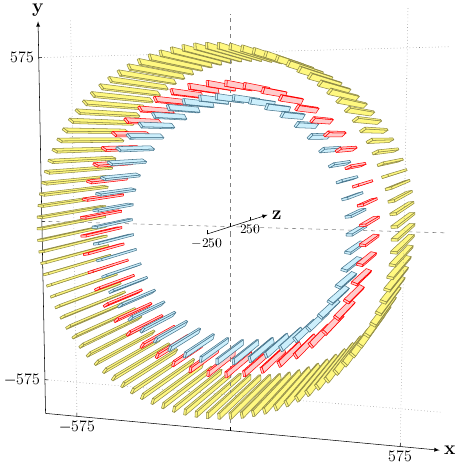}
    \end{subfigure}
    \caption{(left) Photograph of the J-PET detector consisting of 192 plastic scintillators arranged in three layers together with a plastic annihilation chamber containing a $^{22}$Na source. (right) Schematic view of the J-PET detector with a reference frame used in the data analysis and simulations.}
    \label{Fig:JPET}
\end{figure}

The J-PET is designed for medical imaging~\cite{MOS16A} and for the studies of properties of positronium atoms in porous matter and in living organisms~\cite{MOS19A, MOS19N}. It is the first PET-scanner built out of 192 strips of organic scintillators (EJ-230) arranged axially in three concentric cylindrical layers as shown in Fig. \ref{Fig:JPET}. 
The dimension of each scintillator is $1.9~cm~\times~0.7~cm$ $~\times~50.0~cm$ with longest side of scintillator arranged along z-axis as shown in right side of Fig.~\ref{Fig:JPET}. The first and second layer consists of 48 scintillators and the third layer is made of such 96 scintillators. Hamamatsu R9800 vacuum tube photomultipliers are placed at opposite ends of each strip connected optically and converting the scintillation light into electrical signals which are read out by multi-threshold digital electronics~\cite{JINST2017} and the data are collected by trigger-less and re-configurable data acquisition system~\cite{ACTA2017Grzesiek, IEEE}. The read out chain consists of Front-End Electronics (FEE) such as Time-to-Digital or Analog-to-Digital Converters (TDCs or ADCs), data collectors and storage. To process and analyse the data from the experiment and simulation, a dedicated offline framework have been developed ~\cite{KRZ15A, KRZ15B} which is a highly flexible, ROOT-based software package which aids the reconstruction and calibration procedures for the detector. The J-PET detector with large acceptance, high angular resolution and with high timing properties stands out to be well suited for the study of charge conjugation violation in search of decay modes of positronium as shown in Section~\ref{section:my}.


\section{Experimental details}
\label{section:my}
\subsection{Production of positronium using plastic annihilation chamber}
For the production of positronium atoms, a small plastic (PA6) annihilation chamber is placed in the centre of detector with a point like beta-plus radioactive isotope of $^{22}Na$ sandwiched between porous material XAD-4.  Positronium is formed when positrons emitted from the source interacts with the electrons of porous material. It may form in triplet state (with spin S=1, called o-Ps) and singlet state (with spin S=0, called p-Ps), due to charge conjugation o-Ps predominantly annihilates into three photons and p-Ps into 2 photons with an energy of 511 keV. And increment in the formation rate of o-Ps inside the chamber is done by maintaining the vacuum pressure. 


\subsection{Selection of 3$\gamma$ decaying from positronium}
\begin{figure}[htb]
    \begin{subfigure}{}
    \centering
    \includegraphics[width=5.8cm]{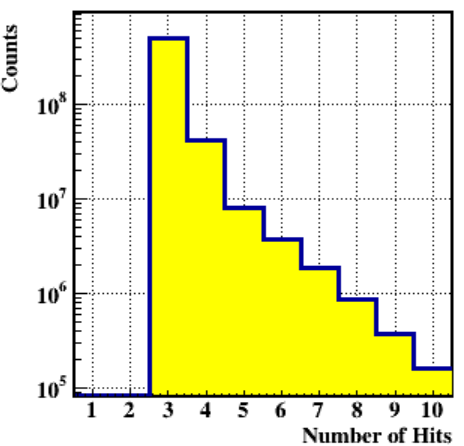}
    \end{subfigure}
    \begin{subfigure}{}
    \includegraphics[width=5.8cm]{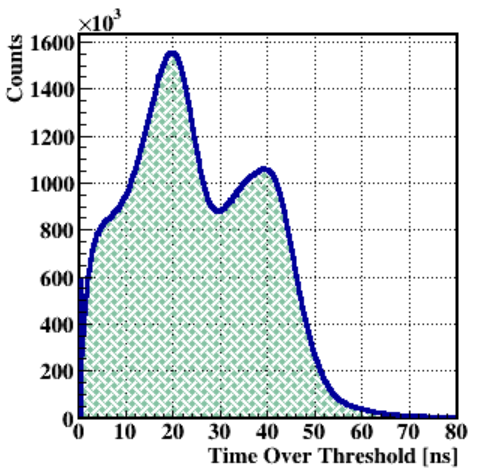}
    \end{subfigure}
    \caption{(left) Experimental distribution representing number of gamma interaction with the detector registered in each event. (right) Experimental TOT distribution representing two Compton spectra from 511 keV and 1274 keV photons.}
    \label{Fig:sumofangles}
\end{figure}

We are interested in registering annihilation of three allowed photons which are decaying from positronium and interacting with the scintillators. Hit-position and hit-time of the $\gamma$ photon in the scintillator can be determined based on the signal instance at both photomultipliers attached to  each  end  of  the  scintillator. In the direction to test C-symmetry, only events in which we are registering three hits which are 3 annihilating photons from the decaying positronium are used. The signal representing multiplicity distribution of hits are presented in Fig.~\ref{Fig:sumofangles}~(left) and only three hit events are considered. Photons are interacting via Compton scattering in the detector since J-PET scanner is built out of plastic scintillators and therefore based on the timing information we estimate the energy deposition of photons using a method called time-over-threshold (TOT)~\cite{JINST2017}. Using TOT distribution (as shown in right side of Fig.~\ref{Fig:sumofangles}) de-excited photons emitted from the source and annihilation photons are separated by making an appropriate cut on the energy.

\begin{figure}[!h]
    \begin{subfigure}{}
    \centering
    \includegraphics[width=5.8cm]{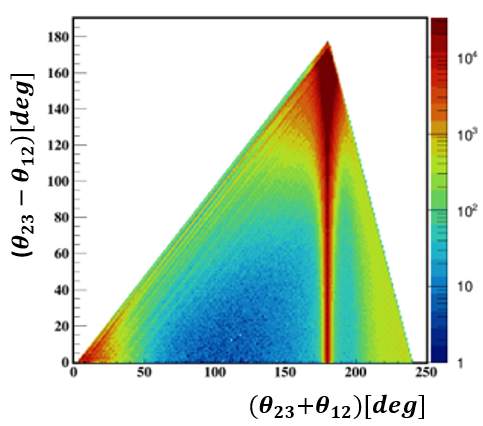}
    \end{subfigure}
    \begin{subfigure}{}
     \includegraphics[width=5.2cm]{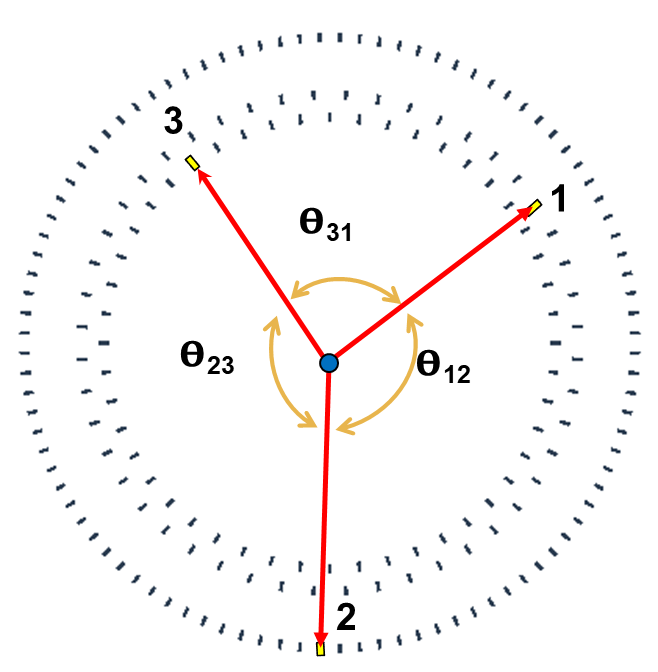}
    \end{subfigure}
    \caption{(left) Experimentally determined relation between the sum and difference of two smallest relative  angles($\Theta_{12}$ and $\Theta_{23}$) of the three interacting annihilation photons and (right) Schematic representation of the simulated 3 photon events.}
    \label{Fig:sumofangles2}
\end{figure}

\begin{figure}[!h]
    \begin{subfigure}{}
    \centering
    \includegraphics[width=5.5cm]{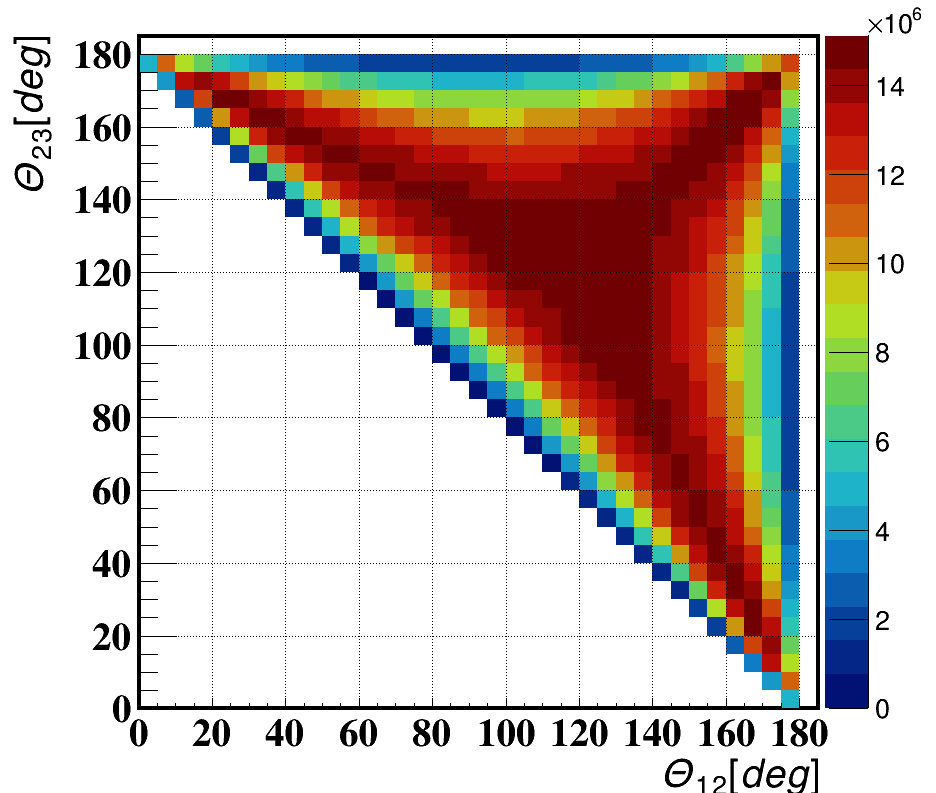}
    \end{subfigure}
    \begin{subfigure}{}
     \includegraphics[width=5.5cm]{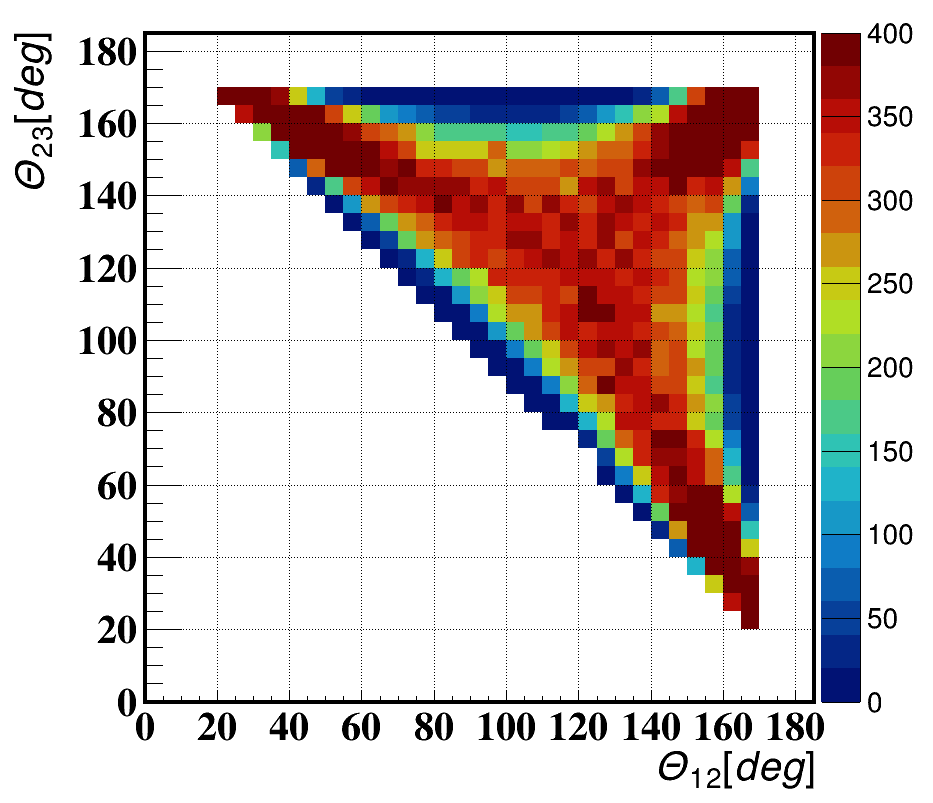}
    \end{subfigure}
    \caption{Distribution of relative angle $\Theta_{12}$ and $\Theta_{23}$ of o-Ps~$\to~3\gamma$ annihilation, for the true generated decays in consideration of relativistic quantum field theory (left) and corresponding reconstructed simulations (right). Boundaries are determined by kinematics constraints.}
    \label{fig:result1}
\end{figure}

Due to charge conjugation conservation, o-Ps predominately annihilates into the three photons and which we refer to as allowed decay. Based on hit position and hit time of each event, momentum direction of each photon is calculated and based on the momenta, relative azimuthal angles between the three photons are determined ~\cite{RAC14A, RAC15A, MOS14A, MOS15A, SHA15A}. According to kinematics (to conserve momentum) we infer that the sum of the two smallest relative azimuthal angles between the registered annihilation photons for o-Ps$~\to~$3$\gamma$ must be greater than 180$^\circ$. Therefore, by making the cut on the spectra as shown in the left side of Fig.~\ref{Fig:sumofangles2}, only the events containing the allowed three photon are considered. Using these relative angles between annihilation photons, as depicted pictorially in the right side of Fig.~\ref{Fig:sumofangles2} it is planned to obtain the angular distribution from experimental data similar to what we obtain from simulated data as shown in Fig.~\ref{fig:result1}.

\newpage
\section{Conclusions and perspectives}
In this article, we presented that the J-PET is an efficient device to measure the o-Ps decay especially, including the charge symmetry violation study. From the simulated data as shown in Fig.~\ref{fig:result1} (left), distribution of relative angles for true generated events are obtained according to relativistic quantum field theory ~\cite{Kamiska2016} for o-Ps decay. The right side of Fig.~\ref{fig:result1} shows distribution of relative angles for registered events from the simulated data taking into account response of J-PET detector geometry. From the analysis of the detector we have shown that the J-PET detector covers almost whole phase space. It is now planned to obtain the angular distribution of allowed photons experimentally, and to separate these three allowed photons (o-Ps~$\to$3~$\gamma$) decay from forbidden three photons (p-Ps~$\to$3$~\gamma$) decay on the basis of lifetime positronium spectra~\cite{KAM18A}.
\section*{Acknowledgement}
This work was supported by the National Science Center of Poland, through OPUS 11 with grant No.~2016/21/B/ST2/01222, and the Foundation of Polish Science through the TEAM POIR.04.04.00-00-4204/17 programme.

\end{document}